\documentclass[journal]{IEEEtran}

\usepackage{amsmath,amsfonts}
\usepackage{algorithmic}
\usepackage{algorithm}
\usepackage{array}
\usepackage{subcaption}
\usepackage{ragged2e, microtype}
\usepackage{pifont}
\usepackage{textcomp}
\usepackage{stfloats}
\usepackage{url}
\usepackage{verbatim}
\usepackage{graphicx}
\usepackage{cite}
\usepackage{makecell}
\usepackage{xcolor}
\usepackage{hyperref}
\usepackage{tablefootnote}
\usepackage{scalerel}
\usepackage{tikz}
\usetikzlibrary{svg.path}
\usepackage{multirow}

\definecolor{orcidlogocol}{HTML}{A6CE39}
\tikzset{
  orcidlogo/.pic={
    \fill[orcidlogocol] svg{M256,128c0,70.7-57.3,128-128,128C57.3,256,0,198.7,0,128C0,57.3,57.3,0,128,0C198.7,0,256,57.3,256,128z};
    \fill[white] svg{M86.3,186.2H70.9V79.1h15.4v48.4V186.2z}
                 svg{M108.9,79.1h41.6c39.6,0,57,28.3,57,53.6c0,27.5-21.5,53.6-56.8,53.6h-41.8V79.1z M124.3,172.4h24.5c34.9,0,42.9-26.5,42.9-39.7c0-21.5-13.7-39.7-43.7-39.7h-23.7V172.4z}
                 svg{M88.7,56.8c0,5.5-4.5,10.1-10.1,10.1c-5.6,0-10.1-4.6-10.1-10.1c0-5.6,4.5-10.1,10.1-10.1C84.2,46.7,88.7,51.3,88.7,56.8z};
  }
}

\newcommand\orcidicon[1]{\href{https://orcid.org/#1}{\mbox{\scalerel*{
\begin{tikzpicture}[yscale=-1,transform shape]
\pic{orcidlogo};
\end{tikzpicture}
}{|}}}}

\newcommand{\notick}{{\color{red}\ding{55}}}

\begin{document}

\title{Adversarial attacks and defenses on ML- and hardware-based IoT device fingerprinting and identification}

\author{Pedro Miguel S\'anchez S\'anchez$^{1}$\orcidicon{0000-0002-6444-2102},
Alberto Huertas Celdr\'an$^{*2}$\orcidicon{0000-0001-7125-1710},
G\'er\^ome Bovet$^{3}$\orcidicon{0000-0002-4534-3483},
Gregorio Mart\'inez P\'erez$^{1}$\orcidicon{0000-0001-5532-6604}

\thanks{$^{*}$Corresponding author.}

\thanks{$^{1}$Pedro Miguel S\'anchez S\'anchez and Gregorio Mart\'inez P\'erez are with the Department of Information and Communications Engineering, University of Murcia, 30100 Murcia, Spain {\tt\small (pedromiguel.sanchez@um.es; gregorio@um.es)}.}%
\thanks{$^{2}$Alberto Huertas Celdr\'an is with the Communication Systems Group (CSG) at the Department of Informatics (IfI), University of Zurich UZH, 8050 Zürich, Switzerland {\tt\small (huertas@ifi.uzh.ch}).}
\thanks{$^{3}$G\'{e}r\^{o}me Bovet is with the Cyber-Defence Campus within armasuisse Science \& Technology, 3602 Thun, Switzerland {\tt\small (gerome.bovet@armasuisse.ch)}.}
\thanks{This work has been partially supported by \textit{(a)} the Swiss Federal Office for Defense Procurement (armasuisse) with the CyberTracer and RESERVE (CYD-C-2020003) projects and \textit{(b)} the University of Zürich UZH.}
}

\markboth{Preprint submitted to IEEE Transactions on Information Forensics and Security}%
{Sanchez \MakeLowercase{\textit{et al.}}: Adversarial attacks and defenses on ML- and hardware-based device fingerprinting and identification}

\maketitle

\begin{abstract}

In the last years, the number of IoT devices deployed has suffered an undoubted explosion, reaching the scale of billions. However, some new cybersecurity issues have appeared together with this development. Some of these issues are the deployment of unauthorized devices, malicious code modification, malware deployment, or vulnerability exploitation. This fact has motivated the requirement for new device identification mechanisms based on behavior monitoring. Besides, these solutions have recently leveraged Machine and Deep Learning techniques due to the advances in this field and the increase in processing capabilities. In contrast, attackers do not stay stalled and have developed adversarial attacks focused on context modification and ML/DL evaluation evasion applied to IoT device identification solutions. However, literature has not yet analyzed in detail the impact of these attacks on individual identification solutions and their countermeasures. This work explores the performance of hardware behavior-based individual device identification, how it is affected by possible context- and ML/DL-focused attacks, and how its resilience can be improved using defense techniques. In this sense, it proposes an LSTM-CNN architecture based on hardware performance behavior for individual device identification. Then, the most usual ML/DL classification techniques have been compared with the proposed architecture using a hardware performance dataset collected from 45 Raspberry Pi devices running identical software. The LSTM-CNN improves previous solutions achieving a +0.96 average F1-Score and 0.8 minimum TPR for all devices. Afterward, context- and ML/DL-focused adversarial attacks were applied against the previous model to test its robustness. A temperature-based context attack was not able to disrupt the identification. However, some ML/DL state-of-the-art evasion attacks were successful. Finally, adversarial training and model distillation defense techniques are selected to improve the model resilience to evasion attacks, improving its robustness without degrading its performance in an impactful manner.

\end{abstract}


\IEEEpeerreviewmaketitle

\section{Introduction}

\IEEEPARstart{T}{he advances} in processing and communication technologies achieved in recent years, enabled by more powerful chips and enhanced connectivity, have motivated an explosion in the deployment of Internet-of-Things (IoT) devices \cite{wang2021evolution}. These devices have generated various use cases and scenarios \cite{shafique2020internet}, such as Industry 4.0, Smart Cities and Homes, or Healthcare. Therefore, the typology of IoT devices is also very heterogeneous depending on the required capabilities of each scenario. In this context, Single-Board Computers (SBC), such as Raspberry Pi, have gained prominence due to their flexibility, relatively high processing power and reduced cost, usually even cheaper than less powerful devices because they are largely manufactured.

However, this increase in processing power not only comes with advantages. Cybersecurity issues have a greater impact when more powerful IoT devices are compromised, as they can perform stronger attacks such as more petitions in a Distributed Denial of Service (DDoS) or calculations for cryptojacking \cite{omolara2022internet}. Therefore, the securitization of the IoT scenario leveraging SBCs is a key factor in guaranteeing its correct functioning. One of the most important aspects is the correct identification of each device deployed, avoiding the presence of unauthorized devices. Static identifiers were traditionally assigned to the devices, but attackers can easily modify and replicate these. To solve this issue, the literature has widely explored the usage of device behavior to identify the devices deployed \cite{sanchez2020survey}.

Behavior-based IoT device identification tasks can be tackled from different granularity levels depending on the environment requirements \cite{sanchez2020survey}. There exist two main approaches: \textit{device model or type identification} (e.g., camera, light bulb, etc.), based on characteristics such as network activities or running processes, and \textit{individual device identification}, where devices from the same model are differentiated based on hardware manufacturing variations using low-level component analysis or radio frequency fingerprinting. Individual device identification is the one offering the best security guarantees. However, it requires lower-level behavior monitoring, as chip manufacturing variations must be analyzed to differentiate devices with the same hardware and software \cite{salo2007multi}. In this sense, hardware performance analysis is one of the most exploited techniques, monitoring how components such as CPU, GPU, or RAM behave or perform when executing a certain task \cite{Sanchez2021methodology}. However, in these solutions, an attacker might exploit the component usage or device context to modify the values generating the fingerprint for device identification and interrupt the identification process.

In the same context and following the trend in other fields, the application of Machine Learning (ML) and Deep Learning (DL) techniques has gained prominence in IoT security during the last years \cite{li2021deep}, including the device identification task \cite{liu2021machine}. But with the deployment of these techniques, adversarial attacks against ML/DL models have appeared \cite{szegedy2013intriguing}, trying to affect the training process or fool the model predictions \cite{sadeghi2020system}. These attacks can affect various factors in the ML/DL pipeline. Usually, they target the training data to poison the model or include backdoors in it \cite{suya2021model}, the testing data to find vulnerabilities in the trained model and change its predictions \cite{kwon2018multi}, or privacy to infer data from the model and its gradients \cite{al2019privacy}. 

ML/DL-based IoT identification solutions have recently been an objective for adversarial attacks \cite{ibitoye2019threat}, demonstrating that these solutions are also affected by context modifications \cite{laor2022drawnapart} or when crafted adversarial samples are employed in the identification process \cite{bao2021threat,namvar2021evaluating}. However, several challenges remain when joining hardware behavior-based individual device identification, ML/DL techniques, and adversarial attacks. Some of these challenges are: (i) which is the best ML/DL technique for device identification based on hardware performance?; (ii) which is the threat model faced by these solutions?; (iii) how device context affects the identification performance?; (iv) how ML/DL-focused adversarial attacks affect the identification process?; (v) how defense techniques improve the resilience to context- and ML/DL-focused adversarial attacks?

To tackle the previous challenges, the main contributions of the present work are:
\begin{itemize}
    
    \item The threat model definition of the adversarial situations that might affect a hardware behavior-based individual device identification solution. It encompasses the complete data lifecycle, from the fingerprint generation to its evaluation, usually using ML/DL techniques.

    \item A LSTM-CNN neural network architecture for individual device identification based on hardware performance behavior of device CPU, GPU, memory and storage. This model considers performance measurements as data points in a time series for data processing and pattern extraction.
    
    \item The architecture performance is compared with different ML/DL classification approaches using the LwHBench dataset \cite{sanchez2022lwhbench}. This dataset contains hardware performance and behavior data from 45 Raspberry Pi devices running identical software images. The proposed LSTM-CNN architecture achieves an average F1-Score of +0.96, correctly identifying all the devices with a +0.80 True Positive Rate. 

    \item The analysis of the impact of different context- and ML/DL-focused adversarial evasion attacks on the individual identification framework. In terms of context attacks, this analysis shows that temperature does not have a big impact on the performance of the identification solution, and that other context conditions such as kernel interruptions can be mitigated during data collection. Regarding ML/DL evasion attacks, the state-of-the-art approaches are successful when performing a targeted attack during evaluation, achieving up to 0.88 attack success rate, and performing a successful device spoofing.

    \item The evaluation of different defense techniques aiming to improve the model robustness against the applied adversarial attacks. The defense techniques applied are adversarial training and model distillation. The results show that the combination of both techniques reduces the attack impact to $\approx$0.18 success rate in the worst case. Additionally, state-of-the-art robustness metrics such as empirical robustness or loss sensitivity also show an effective increase.
    
\end{itemize}

The remainder of this article is structured as follows. Section \ref{sec:related} gives an overview of hardware-based individual device identification, context-focused attacks, and ML/DL-focused attacks. Section \ref{sec:threat_model} gives an overview of the threat model that an IoT device identification solution suffers. Section \ref{sec:identification} describes the ML- and hardware-based device fingerprinting solution for individual device identification. Section \ref{sec:attacks} gives an overview of the implementation and impact of the adversarial attack on device identification. Next, Section \ref{sec:defenses} describes how the application of defense mechanisms enhances the solution resilience against adversarial attacks. Finally, Section \ref{sec:conclusion} gives an overview of the conclusions extracted from the present work and future research directions.

\section{Related Work}
\label{sec:related}

This section gives the insights required to understand the concepts used in the following sections and reviews the main works in the literature associated with the present one.

\subsection{Hardware-based individual device identification}

The present work focuses on hardware-based individual device identification using the behavior and performance of the components contained in the device. In \cite{Sanchez2021methodology}, the authors compared the deviation between the CPU and GPU cycle counters in Raspberry Pi devices to perform individual identification of 25 devices. The identification was performed using XGBoost, achieving a 91.92\% True Positive Rate (TPR). Similarly, \cite{laor2022drawnapart} performed identical device identification using GPU performance behavior and ML/DL classification algorithms. Accuracy between 95.8\% and 32.7\% was achieved in nine sets of identical devices, including computers and mobile devices.

Sanchez-Rola et al. \cite{Sanchez2018clock} identified +260 identical computers by measuring the differences in code execution performance. They employed the Real-Time Clock (RTC), which includes its own physical oscillator, to find slight variations in the performance of each CPU. In \cite{salo2007multi}, the author compared the drift between the CPU time counter, the RTC chip, and the sound card Digital Signal Processor (DSP) to identify identical computers. 

Other works have explored the usage of Physical Unclonable Functions (PUFs) for IoT device identification \cite{shamsoshoara2020survey}. However, PUFs are out of the scope of this work, as it is centered on hardware behavior fingerprinting based on device performance.

\subsection{Context-focused attacks}

In hardware-based identification solutions, the context in which the identification code or tasks are executed might influence the collected data and, therefore, the results achieved. In this sense, the temperature can affect the frequency of crystal oscillators or hardware load might introduce delays due to the scheduling between processes. Therefore, a malicious attacker could change these context conditions to affect the identification, making it unusable or generating measurements that mimic another device.

The works described in the previous section briefly discussed context issues that may affect the identification process. \cite{Sanchez2021methodology} showed that device rebooting and other processes running in the device impacted the identification results if proper process isolation mechanisms for data collection were not implemented. Besides, they checked that usual temperature changes based on device load did not affect the results. \cite{laor2022drawnapart} demonstrated that environment temperatures between 26.4ºC and 37ºC did not affected to the identification results. However, rebooting had an impact on the identification, dropping the results to 50.3\% accuracy. The authors also leave voltage variations as a future line to evaluate. In \cite{Sanchez2018clock}, the authors evaluated the identification application under different CPU loads and temperatures, with positive results in both cases. Finally, \cite{salo2007multi} only mentioned temperature impact analysis as part of future work and no context-based experiments were performed.

As can be seen, none of the previous works on device fingerprinting and identification based on hardware performance behavior has extensively explored the impact that context-focused attacks may have on their results.

\subsection{ML/DL-focused adversarial attacks}

Adversarial ML/DL \cite{sadeghi2020system} is a research field that seeks to develop not only accurate models, but also highly robust models against tampering. It studies the possible attacks against ML/DL models as well as the defense techniques that can secure these. There exist a wide variety of taxonomies for the attacks that an ML/DL model may suffer. In this sense, attacks can be classified in: (i) \textit{Poisoning}, when the model is attacked during training using malicious samples; (ii) \textit{Evasion}, when the model evaluation process is attacked, trying to fool a legitimately trained model; and (iii) \textit{Model Extraction} attacks, where the attacker tries to infer the model based on its predictions.

In this work, the focus is on evasion attacks, as the intention is to full a model trained for device identification, making it misclassify a malicious device for the legitimate one. Here, the main types of attacks are: \textit{non-targeted attacks}, when the objective is just to misclassify a sample to any different class that is not the original one, and \textit{targeted attacks}, when the objective is to evaluate the malicious sample as a concrete objective class. Several evasion attacks can be found as the most common ones in the literature:
\begin{itemize}
    
\item As one of the first evasion attacks for DL, Goodfellow et al. \cite{goodfellow2014explaining} proposed the Fast Gradient Sign Method (FGSM). This attack performs one-step updates in the adversarial sample following the direction of the gradient loss, trying to move the sample into the boundary of a different class. The equation characterizing FGSM can be seen as: 

\begin{equation}
    X_{adv}=X + \epsilon*sign(\nabla_{x}J(X,Y))
\end{equation}

where $\epsilon$ is a parameter defining the size of the perturbation update and $\nabla_{x}J(X,Y)$ is the gradient loss function for the sample $X$.

\item Basic Iterative Method (BIM) \cite{kurakin2018adversarial} is a improvement over FGSM by including iterative optimization. This is, applying FGSM several times with small perturbation steps.

\item Momentum Iterative Method (MIM) \cite{dong2018boosting} integrates momentum into the iterative FGSM or BIM, avoiding local minimum or overfitting influence in the generated adversarial samples.

\item Projected Gradient Descent (PGD) \cite{madry2017towards} is a generalization of BIM that has no constraints on the iteration steps. 

\item DeepFool $L_2$ attack \cite{moosavi2016deepfool} minimizes the Euclidean distance between the original and the adversarial samples by estimating the model decision boundary using a linear classifier.

\item Jacobian-based Saliency Map Attack (JSMA) \cite{papernot2016limitations} is another common attack in the literature. It uses the Jacobian matrix \cite{waldron1985study} of the model to find the sensitivity direction of the model and perform feature selection to minimize the number of characteristics modified from the original data sample.

\item Boundary Attack \cite{brendel2017decision} generates a random adversarial sample and then performs optimizations in the $L_2-norm$ of the perturbation to make the sample similar to the original legitimate vector, but maintaining the misclassification result.

\item Carlini\&Wagner (C\&W) Attack \cite{carlini2017towards} proposes a optimization-based adversarial sample generation. It can be applied to three distance metrics: $L_0, L_2, L_\infty$. $L_0$ measures the number of features to be modified, $L_2$ measures the Euclidean distance between a benign and adversarial sample, and $L_\infty$ measures the maximum change to any feature.

\item Generative Adversarial Network (GAN)-based attack \cite{hu2017generating} uses GAN models to generate realistic adversarial samples able to fool the classifier.

\end{itemize}

Numerous defense mechanisms have arisen against the previous attacks and others that may be found in the literature \cite{rosenberg2021adversarial}. The objective of these countermeasurements is to make the models resistant to adversarial samples. Generally, these can be classified into detection and robustness methods, depending if the aim is to detect crafted malicious samples prior to evaluation or make the model resistant to the evaluation of these, respectively. Besides, defense mechanisms can be attack-specific or attack-agnostic, depending on whether they are focused on improving resilience against a specific attack.

One of the most extended defense techniques to avoid evasion attacks is Adversarial Training \cite{wong2020fast}, where malicious samples are employed for model training, avoiding the impact of the attacks that generated those samples. Knowledge distillation \cite{hinton2015distilling} has also been applied for robustness improvement at training. This technique seeks to generate smaller models using the base model outputs as features. It can improve the model robustness by generating smoother decision boundaries and less sensitivity to adversarial samples \cite{papernot2016distillation}.

\subsubsection{Adversarial ML in IoT identification and security}

In this sense, \cite{bao2021threat} is the closest work to the one at hand. The authors analyzed the impact of different non-targeted and targeted adversarial attacks (FGSM, BIM, PGD and MIM) over a CNN implemented for radiofrequency-based individual device identification. Similarly, Namvar et al. \cite{namvar2021evaluating} evaluated the resilience of network-based IoT identification ML solutions against adversarial samples generated with FGSM, BIM, and JSMA. They showed how classifier models giving +90\% accuracy decrease their performance to 75-55\% when exposed to maliciously crafted samples. From a different perspective, Benegui and Ionescu \cite{benegui2020adversarial} evaluated the impact of adversarial samples over ML/DL models for user identification based on motion sensors, achieving near to 100\% attack success rates with FGSM, JSMA, DeepFool, and Boundary Attacks. Later, \cite{pourshahrokhi2021generative} demonstrated that a GAN-based attack has more impact than the previous attacks in the user identification context.

From a more generic perspective, \cite{ibitoye2019threat} and \cite{apruzzese2022modeling} reviewed the threat of adversarial attacks in ML solutions applied in network security. They proved the high impact of adversarial attacks over ML-based security systems, highlighting the need for more research on attack and defense methods in the area.

\tablename~\ref{tab:related} shows a comparison between the different solutions reviewed in this section. It can be seen how none of the previous works combines the application of context- and ML/DL-focused attacks. Besides, the ML/DL-focused attack papers in the context of device or user identification have not explored the reward from the defense mechanisms available in the literature. Therefore, the present work solves a gap in the literature, providing useful insights in the impact of attack and defense techniques on the context of hardware- and ML/DL-based individual device identification. 

\begin{table*}[]
\centering
\scriptsize
\caption{Comparison of previous works on Context and ML/DL-focused adversarial attacks in identification solutions}
\label{tab:related}
\begin{tabular}{>{\centering\arraybackslash}m{0.65cm}>{\centering\arraybackslash}m{1.5cm}>{\centering\arraybackslash}m{2cm}>{\raggedright\arraybackslash}m{4cm}>{\centering\arraybackslash}m{2.0cm}>{\raggedright\arraybackslash}m{5.5cm}}

\textbf{Work} & \makecell[c]{\textbf{Scenario}}& \makecell[c]{\textbf{Attack Type}} & \textbf{\makecell[c]{Attack Technique}}& \textbf{\makecell[c]{Defense}} & \textbf{\makecell[c]{Results}} \\ \hline  
\hline
\cite{salo2007multi} (2007) & Computer identification & \makecell[c]{\notick} & \makecell[c]{\notick} & \notick & Computer identification based on the comparison of three physical oscillators using t-test statistic \\
\hline
\cite{Sanchez2018clock} (2018) & Computer identification & Context-focused & CPU load, temperature & Process isolation & 265 computers uniquely identified. No effect from CPU load and temperature \\
\hline
\cite{Sanchez2021methodology} (2021) & IoT device identification & Context-focused & Temperature changes and device rebooting & Process isolation & 91.92\% average TPR in 25 devices. No effects from temperature changes and device rebooting \\
\hline
\cite{laor2022drawnapart} (2022) & Computer and mobile identification & Context-focused & Temperature changes and device reboot & \notick & 95.8\% and 32.7\% accuracy in nine sets of identical devices. Accuracy drop with device rebooting \\
\hline
\hline
\cite{benegui2020adversarial} (2020) & User identification & DL-focused & Non-targeted and targeted attacks (FGSM, JSMA, DeepFool, Boundary) & \makecell[c]{\notick} & Near to 100\% attack success over CNNs models with different depths (from 4 to 12 layers) \\
\hline
\cite{pourshahrokhi2021generative} (2021) & User identification & DL-focused & GAN-based attack & \makecell[c]{\notick} & GAN-generated samples were more effective than FGSM, Deepfool and Boundary when performing adversarial attacks \\
\hline
\cite{namvar2021evaluating} (2021) & IoT device identification& ML/DL-focused & Non-targeted attacks (FGSM, BIM, JSMA) & \makecell[c]{\notick} & Accuracy decreased from +90\% to 75-55\% \\
\hline
\cite{bao2021threat} (2021) & IoT device identification & ML/DL-focused & Non-targeted and targeted attacks (FGSM, BIM, PGD and MIM) & \makecell[c]{\notick} & Proven vulnerability to targeted attacks with +80\% attack success rate. \\
\hline
\hline
\textbf{This work (2022)} & IoT device identification & Context and ML/DL-focused & Context: Temperature changes, CPU load, device rebooting \newline ML/DL: FGSM, BIM, MIM, PGD, JSMA, Boundary Attack, C\&W  & Process isolation, Adversarial training, Model distillation & +0.96 average F1-Score. Resilience to temperature and process-based context attacks. ML/DL evasion attack resilience improved using model distillation and adversarial training. \\
\hline
\end{tabular}%
\end{table*}

\section{Threat Model}
\label{sec:threat_model}

This section details the threat model faced by an ML/DL-based device identification solution based on hardware performance. Note that this threat model focuses on the moment when the identification solution is already trained and deployed, so only threats affecting device evaluation are considered. In this sense, an attacker may try to affect the two different sides of the identification: (i) the hardware generating the data or (ii) the ML/DL models in charge of the data evaluation. Considering that, the following threats have been identified:

\begin{itemize}
    \item \textit{TH1. Fingerprint eavesdropping and hijacking}. An adversary could read the data composing a fingerprint, either at the level of in-device data collection, communication, or during processing (in a server or the device itself), and then use it in another device to impersonate the identity of the first. This threat implies a reduced knowledge of the fingerprint generation process and the functions and components used during the process.

    
    \item \textit{TH2. Fingerprint forgery}. Since the components and frequencies of the devices are public, an attacker with knowledge about the functions that are executed to generate the fingerprint could try to generate a new one that resembles that of a legitimate device. This threat would be triggered possibly on a trial/error or brute force basis. This threat requires thorough knowledge of the implementation of the fingerprint generation process and the values composing the fingerprint.

    \item \textit{TH3. Context modification}. As the fingerprint is based on data collected from the performance of the execution of certain tasks in the software, an attacker may try to modify the conditions under which the fingerprint is generated. This can neglect to successfully recognize a legitimate device or generate fingerprints that pretend to mimic another device. The context can be modified from several perspectives, for example raising the device temperature (using external tools or exhaustively using the hardware) or introducing software that may add kernel interruptions in the fingerprint collection program, which should be isolated from these interactions as much as possible. 

    \item \textit{TH4. ML/DL evaluation evasion}. In ML/DL-based solutions, an attacker with enough knowledge or access to the evaluation model can be able to craft malicious data samples to fool the ML/DL solution. These samples can target and impersonate a specific device following a trial and error approach or using a targeted attack. In this sense, several adversarial attacks have been proposed in the literature as shown in Section \ref{sec:related}.
    
\end{itemize}

Therefore, a proper individual device identification solution has to consider and evaluate the previous threats in order to ensure correct functioning and attack resilience.

\section{Individual Device Identification}
\label{sec:identification}

The present section describes the ML/DL framework implemented for hardware-based individual device identification. It sets the baseline results for later attack and defense technique impact analysis.

\subsection{Dataset collection and preprocessing}

\subsubsection{Dataset collection} For individual device identification based on hardware behavior, the imperfections in the chips contained in the device should be monitored to be later evaluated. As seen in Section \ref{sec:related}, this task has usually been tackled in the literature by comparing components using different crystal oscillators or base frequencies, as deviations in the performance of these components can be noticed from the device itself.

To implement the individual device identification framework, a dataset leveraging metrics related to the hardware components contained in some devices was required. The dataset was named LwHbench and more details are available in \cite{sanchez2022lwhbench}. In this sense, a dataset collected performance metrics from CPU, GPU, Memory, and Storage from 45 Raspberry Pi devices from different models for 100 days. Different functions were executed in these components where other hardware elements (running at different frequencies) were employed for performance measurement. \tablename~\ref{tab:features_dataset} summarizes the set of functions monitored.

\begin{table}[htpb!]
    \centering
    \scriptsize
    \caption{Features available in LwHBench dataset \cite{sanchez2022lwhbench}.}
    \begin{tabular}{ >{\raggedright\arraybackslash}m{1.5cm} >{\raggedright\arraybackslash}m{1.5cm} >{\raggedright\arraybackslash}m{4.55cm} } 
    \textbf{Component} & \textbf{Function} & \textbf{Monitored Feature} \\
    \hline
    \hline
    \textbf{-} & timestamp & Unix timestamp \\
    & temperature & Device core temperature \\
    \hline
    \textbf{CPU} & 1 s sleep & GPU cycles elapsed during 1 second CPU sleep \\
     & 2 s sleep & GPU cycles elapsed during 2 seconds CPU sleep \\
     & 5 s sleep & GPU cycles elapsed during 5 seconds CPU sleep \\
     & 10 s sleep & GPU cycles elapsed during 10 seconds CPU sleep \\
     & 120 s sleep & GPU cycles elapsed during 120 seconds CPU sleep \\
     & string hash & GPU cycles elapsed during a fixed string hash calculation \\
     & pseudo random & GPU cycles elapsed while generating a software pseudo-random number \\
     & urandom & GPU cycles elapsed while generating 100 MB using \textit{/dev/urandom} interface \\
     & fib & GPU cycles elapsed while calculating Fibonacci number for 20 using the CPU \\
     \hline
    \textbf{GPU} & matrix mul & CPU time taken to execute a GPU-based matrix multiplication \\
     & matrix sum & CPU time taken to execute a GPU-based matrix summation \\
     & scopy & CPU time taken to execute a GPU-based graph shadow processing \\
    \hline
     \textbf{Memory} & list creation & CPU time taken to generate a list with 1000 elements \\
     & mem reserve & CPU time taken to fill 100 MB in memory \\
     & csv read & CPU time taken to read a 500 kB \textit{csv} file \\
    \hline
    \textbf{Storage} & read x100 & 100 CPU time measurements for 100 kB storage read operations \\
     & write x100 & 100 CPU time measurements for 100 kB storage write operations \\
    \hline
    \end{tabular}
    \label{tab:features_dataset}
\end{table}

The dataset contains per device model: 505584 samples from 10 RPi 1B+, 784095 samples from 15 RPi4, 547800 samples from 10 RPi3 and 548647 samples from 10 RPiZero. For data collection, several countermeasures were taken to avoid the effect of noise introduced by other processes running in the devices: Fixed component frequency, kernel level priority, code executed in an isolated CPU core (in multi-core devices) and the disabling of memory address randomization. Besides, the dataset was collected under several temperature conditions, allowing the impact analysis of this context characteristic in the component performance. 

\subsubsection{Data preprocessing} As the first preprocessing technique and following the approach of \cite{Sanchez2021methodology}, sliding-window-based feature extraction was performed per device, extracting statistical features such as median, average, maximum, minimum, and summation. The reasoning behind this preprocessing is that the distribution of raw feature values from each device may overlap due to the limited variability in the component performance. Therefore, statistical values such as median or average help to differentiate between partially overlapping distributions. Only some of the available raw features were selected for this step, as keeping a low feature number helps to lighten the ML/DL model training. \tablename~\ref{tab:features} describes the set of features extracted from the dataset of each device.

\begin{table}[htpb]
	\caption{Feature set extracted for validation.}
	\centering
    \scriptsize
     \begin{tabular}{>{\RaggedRight\arraybackslash}p{1.2cm}>{\RaggedRight\arraybackslash}p{1.90cm}>{\RaggedRight\arraybackslash}p{1.5cm}
    >{\RaggedRight\arraybackslash}p{1.1cm}c}
        \hline
        \textbf{\makecell[c]{Operation\\Collected}} & \textbf{\makecell[c]{Python Code\\Function}} & \textbf{\makecell[c]{Sliding\\Windows}} & \textbf{\makecell[c]{Statistics\\Extracted}} & \textbf{\makecell[c]{No.\\Features}}\\
        \hline
        10 s sleep & \textit{time.sleep(10)} & \multicolumn{1}{|c}{}&\multicolumn{1}{|c|}{} & 40\\
        \cline{1-2}     \cline{5-5}
        120 s sleep & \textit{time.sleep(120)} & \multicolumn{1}{|>{\RaggedRight\arraybackslash}p{1.4cm}}{} & \multicolumn{1}{|>{\RaggedRight\arraybackslash}p{1.1cm}|}{ } & 40\\
        \cline{1-2}     \cline{5-5}
        string hashing & \textit{hashlib.sha256(str)} & \multicolumn{1}{|>{\RaggedRight\arraybackslash}p{1.5cm}}{10 Sliding windows.} & \multicolumn{1}{|>{\RaggedRight\arraybackslash}p{1.1cm}|}{} & 40 \\
        \cline{1-2}     \cline{5-5}
         urandom & \textit{os.urandom()} & \multicolumn{1}{|>{\RaggedRight\arraybackslash}p{1.4cm}}{Group sizes: } & \multicolumn{1}{|>{\RaggedRight\arraybackslash}p{1.1cm}|}{Minimum,} & 40\\
         \cline{1-2}     \cline{5-5}

        matrix mul & \textit{vc.cond\_mul()} & \multicolumn{1}{|>{\RaggedRight\arraybackslash}p{1.5cm}}{10, 20, 30, 40,} & \multicolumn{1}{|>{\RaggedRight\arraybackslash}p{1.1cm}|}{maximum,}& 40\\
         \cline{1-2}     \cline{5-5}
        matrix sum & \textit{vc.cond\_add()} & \multicolumn{1}{|>{\RaggedRight\arraybackslash}p{1.4cm}}{50, 60, 70,} & \multicolumn{1}{|>{\RaggedRight\arraybackslash}p{1.1cm}|}{mean,}& 40\\
         \cline{1-2}     \cline{5-5}

        list creation & \textit{list.append()} & \multicolumn{1}{|>{\RaggedRight\arraybackslash}p{1.4cm}}{80, 90, 100} & \multicolumn{1}{|>{\RaggedRight\arraybackslash}p{1.1cm}|}{median}& 40\\
        \cline{1-2}     \cline{5-5}
        memory reserve & \textit{cgroup.set\_memory()} & \multicolumn{1}{|>{\RaggedRight\arraybackslash}p{1.4cm}}{} & \multicolumn{1}{|>{\RaggedRight\arraybackslash}p{1.1cm}|}{}& 40\\
        \cline{1-2}     \cline{5-5}
        
        CSV read & \textit{pandas.read\_csv()} & \multicolumn{1}{|>{\RaggedRight\arraybackslash}p{1.4cm}}{} & \multicolumn{1}{|>{\RaggedRight\arraybackslash}p{1.1cm}|}{}& 40\\
        \cline{1-2}     \cline{5-5}
         
        1\textsuperscript{st} storage read & \textit{os.read()} & \multicolumn{1}{|>{\RaggedRight\arraybackslash}p{1.4cm}}{} & \multicolumn{1}{|>{\RaggedRight\arraybackslash}p{1.1cm}|}{}& 40\\
        \cline{1-2}     \cline{5-5}
        
        1\textsuperscript{st} storage write & \textit{os.read()} & \multicolumn{1}{|>{\RaggedRight\arraybackslash}p{1.4cm}}{} & \multicolumn{1}{|>{\RaggedRight\arraybackslash}p{1.1cm}|}{}& 40\\
        
        \hline
        Total & & & & 440\\
        \hline
    \end{tabular}
    \label{tab:features}
\end{table}{}

In addition to sliding windows, it was decided to directly evaluate the raw data vectors without the sliding window processing described above. The reasoning behind this approach is that having a large dataset of raw values can work well in the case of DL models, which can automatically extract internal insights from the data. In this approach, only timestamp and temperature features were filtered, using the rest of the values (215 values in total) as features for the models. 

Finally, 
it is also decided to perform a time series-based evaluation, concatenating together the available samples in groups of 10 vectors. This grouping technique allows the application of time series DL methods such as LSTM and 1D-CNN models \cite{yu2019review,kiranyaz20211d}. These models can extract complex trends in the data that may achieve better results than the isolated processing and evaluation of individual samples.

\subsection{LSTM-1DCNN architecture}

This work proposes a framework that leverages an LSTM-1DCNN neural network architecture for the classification of the performance samples obtained from the device. These models have shown good performance in very varied time series scenarios, such as human activity recognition \cite{xia2020lstm}, gold price forecast \cite{he2019gold}, or DNA protein binding \cite{zhang2020deepsite}.



The network architecture combines LSTM and 1D-CNN layers to extract patterns in the series fed as input. The main benefit of this approach is that combines the recursion patterns extracted by the LSTM layer, due to its memory capabilities, with the space patterns extracted by the 1D-CNN layer, as kernels are applied to close features to derive more complex ones.

\begin{figure}[htpb!]
    \centering
    \includegraphics[width=\columnwidth,trim={0 0 0 0} ,clip=true]{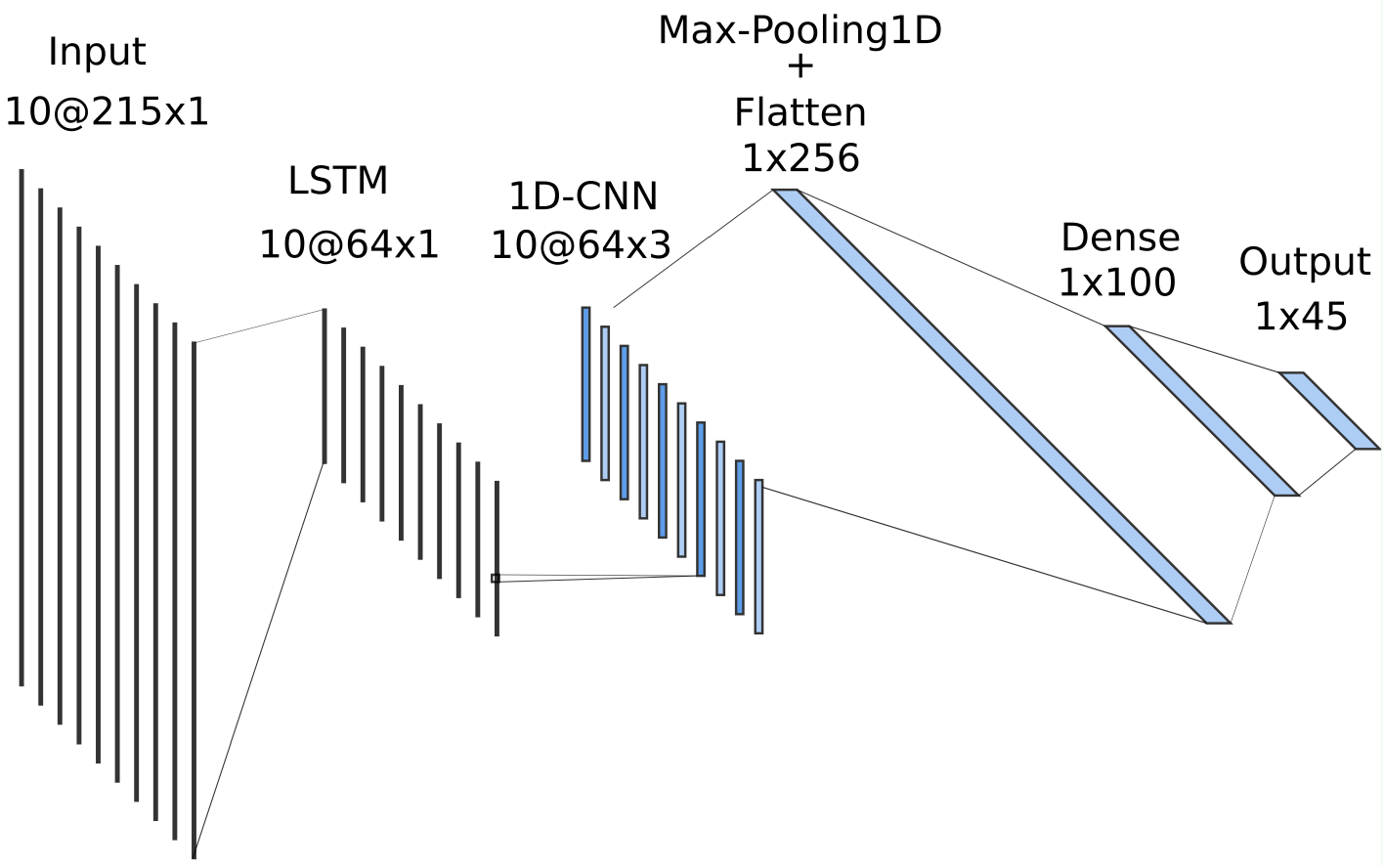}
    \caption{LSTM-1DCNN architecture proposed.}
    \label{fig:nn}
\end{figure}

\figurename~\ref{fig:nn} describes the neural network architecture explained above, depicting the size of each layer. The LSTM layer is configured to return sequences, so the 1D-CNN layer can be applied afterward in those sequences. After the 1D-CNN layer, Max-Pooling is applied. Finally, a fully connected layer of 100 neurons is added before the last layer with 45 outputs, one per device. In the implementation, the LSTM layer has 64 neurons, the 1D-CNN layer uses \textit{ReLU} is used as activation function in the hidden layers and \textit{ADAM} is used as optimizer during training (\tablename~\ref{tab:clasif_alg_hyp} show the complete list of hyperparameters tested).

\subsection{Classification-based device identification experiments}

Once the two data preprocessing approaches were applied to generate two datasets, one with raw values and another with sliding-window-based features, the next step was to compare the proposed LSTM-1DCNN model with the most common ML/DL classification approaches. In \cite{Sanchez2021methodology}, the authors directly applied ML classifiers using CPU and GPU-related statistical features similar to the ones described in the previous section. Moreover, LSTM and 1D-CNN networks were also tested for the time series approaches. Finally, a more complex multi-input network that combined one LSTM and one 1D-CNN input layer was also implemented for comparison, this model is denoted as Multi\_1DCNN\_LSTM. The experiments were performed in a server equipped with a \textit{AMD EPYC 7742} CPU and a \textit{NVIDIA A100} GPU.

\tablename~\ref{tab:clasif_alg_hyp} describes the algorithms and hyperparameters tested. Besides, for the algorithms requiring data normalization, \textit{QuantileTransformer} \cite{ahsan2021effect} was applied, as the data from the different device models had different distributions based on their hardware capabilities. 80\% of the data was used for training and cross-validation, while 20\% was used for testing. The train/test splitting was done without vector shuffling to avoid that possible order correlation in the vectors might influence the results.

\begin{table}[htpb!]
	\caption{Classification algorithms and hyperparameters tested against the proposed architecture.}
	\centering
    \scriptsize
    \begin{tabular}{m{1.1cm}m{6.9cm}}
        \hline
        \textbf{Model} & \makecell[c]{\textbf{Hyperparameters tested}} \\
        \hline
        Naive Bayes & No hyperparameter tunning required  \\
        \hline
        k-NN & $k\in [3,20]$ \\
        \hline
        SVM & \makecell[l]{$C\in [0.01,100], gamma\in [0.001,10]$\\$kernel\in \{'rbf', 'linear', 'sigmoid','poly'\}$} \\
        \hline
        AdaBoost & $n\_estimators \in [10,100]$ \\
        \hline
        XGBoost & \makecell[l]{$lr \in [0.01,0.3], max\_depth \in [3,15]$\\ $min\_child\_weight\in [1,7], gamma \in [0,0.5],$\\$colsample\_bytree \in [0.3,0.7]$} \\
        \hline
        Decision Tree & \makecell[l]{$max\_depth \in [None, 5, 10, 15, 20]$ \\ $min\_samples\_split \in [2,3,4,5]$} \\
        \hline
        \makecell[l]{Random\\Forest} & \makecell[l]{$number\_of\_trees \in [50,1000]$\\$max\_depth \in [None, 5, 10, 15, 20]$\\$min\_samples\_split \in [2,3,4,5]$} \\
        \hline
        MLP & \makecell[l]{$n\_layers\in [1,3], neurons\_layer\in [100,500], $ \\ $batch\_size \in [32,64,128,256,512]$ \\ $activation = relu, optimizer=[SGD, adam, adamax]$} \\
        \hline
        1D-CNN & \makecell[l]{$filters=[16,32,64,128],kernel\_size=[3,5,7],$\\$n\_layers=[1,2,3],optimizer=[SGD, adam, adamax] $}\\
        \hline
        LSTM & \makecell[l]{$neurons=[10,100], n\_layers=[1,2,3],$\\$optimizer=[SGD, adam, adamax]$}\\
        \hline
        \makecell[l]{Multi\_\\1DCNN\_\\LSTM} & \makecell[l]{$input\_layers=[2,3], cnn\_filters=[16,32,64,128],$ \\$cnn\_kernel\_size=[3,5,7], lstm\_neurons=[10,100]$\\$n\_layers=[1,2,3], optimizer=[SGD, adam, adamax]$}\\
        \hline
    \end{tabular}
    \label{tab:clasif_alg_hyp}
\end{table}{}

\tablename~\ref{tab:classif_results} depicts the classification results for each algorithm (with its best hyperparameter setup) in both of the generated datasets. The performance metrics are Accuracy, average Precision, average Recall, and average F1-Score: (TP: True Positives, TN: True Negatives, FP: False Positives, FN: False Negatives)

\begin{equation}
    Accuracy=\frac{TP+TN}{TP+FN+TN+FP}
\end{equation}
\begin{equation}
    Precision=\frac{TP}{TP+FP}
\end{equation}
\begin{equation}
    Recall\ or\ TPR =\frac{TP}{TP+FN}
\end{equation}
\begin{equation}
    F1-Score=\frac{2 \times precision \times recall}{precision + recall}
\end{equation}

\begin{table*}[htpb!]
	\caption{Baseline classification model performance.}
	\centering
    \scriptsize
    \begin{tabular}{m{2.3cm}|cccc|cccc|}
        \cline{2-9}
        \cline{2-9}
        & \multicolumn{4}{c|}{\textbf{Raw data features}} & & \multicolumn{2}{c}{\textbf{Sliding-window features}} & \\
        \hline
        \textbf{Model} & \textbf{Accuracy} & \textbf{Avg. Precision} & \textbf{Avg. Recall} & \textbf{Avg. F1} & \textbf{Accuracy} & \textbf{Avg. Precision} & \textbf{Avg. Recall} & \textbf{Avg. F1} \\
        \hline
        & \multicolumn{8}{c|}{\textbf{Single vector approaches}} \\
        \hline
        Naive Bayes & 0.4569 & 0.4735 & 0.4569 & 0.4473 & 0.6829 & 0.6935 & 0.6829 & 0.6719 \\
        \hline
         k-NN & 0.4526 & 0.4679 & 0.4526 & 0.4472 & 0.5274 & 0.5410 & 0.5274 & 0.5285 \\
         \hline
         SVM & 0.7838 & 0.7955 & 0.7829 & 0.7849 & 0.7375 & 0.7434 & 0.7318 & 0.7297 \\
         \hline
        AdaBoost & 0.0705 & 0.0060 & 0.0705 & 0.0110 & 0.0706 & 0.0060 & 0.0706 & 0.0110 \\
         \hline
         XGBoost & 0.9059 &0.9173 &0.9056 &0.9087 & 0.7498 & 0.7655 & 0.7498 & 0.7461 \\
         \hline
         Decision Tree & 0.7816 & 0.7896 & 0.7825 & 0.7837 & 0.6932 & 0.7045 & 0.6932 & 0.6910  \\
         \hline
         Random Forest & 0.8549 & 0.8664 & 0.8542 & 0.8570 & 0.7487 & 0.7615 & 0.7487 & 0.7430 \\
         \hline
         MLP & 0.8895 & 0.8960 & 0.8880 & 0.8899 & 0.6840 & 0.6988 & 0.6758 & 0.6749 \\
         \hline
         & \multicolumn{8}{c|}{\textbf{Time series approaches (10 values)}} \\
         \hline
        1D-CNN & 0.9428 & 0.9453 & 0.9428 & 0.9428 & 0.6941 & 0.7170 & 0.6941 & 0.6862  \\
        \hline
        LSTM & 0.9346 & 0.9430 & 0.9346 & 0.9346 & 0.7225 & 0.7345 & 0.7225 & 0.7147 \\
        \hline
        LSTM\_1D-CNN & \textbf{0.9602} & \textbf{0.9626} & \textbf{0.9602} & \textbf{0.9602} & 0.7149 & 0.7287 & 0.7149 & 0.7080 \\
        \hline
        Multi\_1DCNN\_LSTM & 0.9535 & 0.9553 & 0.9535 & 0.9535 & 0.6784 & 0.6947 & 0.6784 & 0.6700 \\
        \hline 
        \hline
    \end{tabular}
    \label{tab:classif_results}
\end{table*}{}

\begin{figure*}[htpb!]
    \centering
    \includegraphics[width=\textwidth,trim={0 115 0 0} ,clip=true]{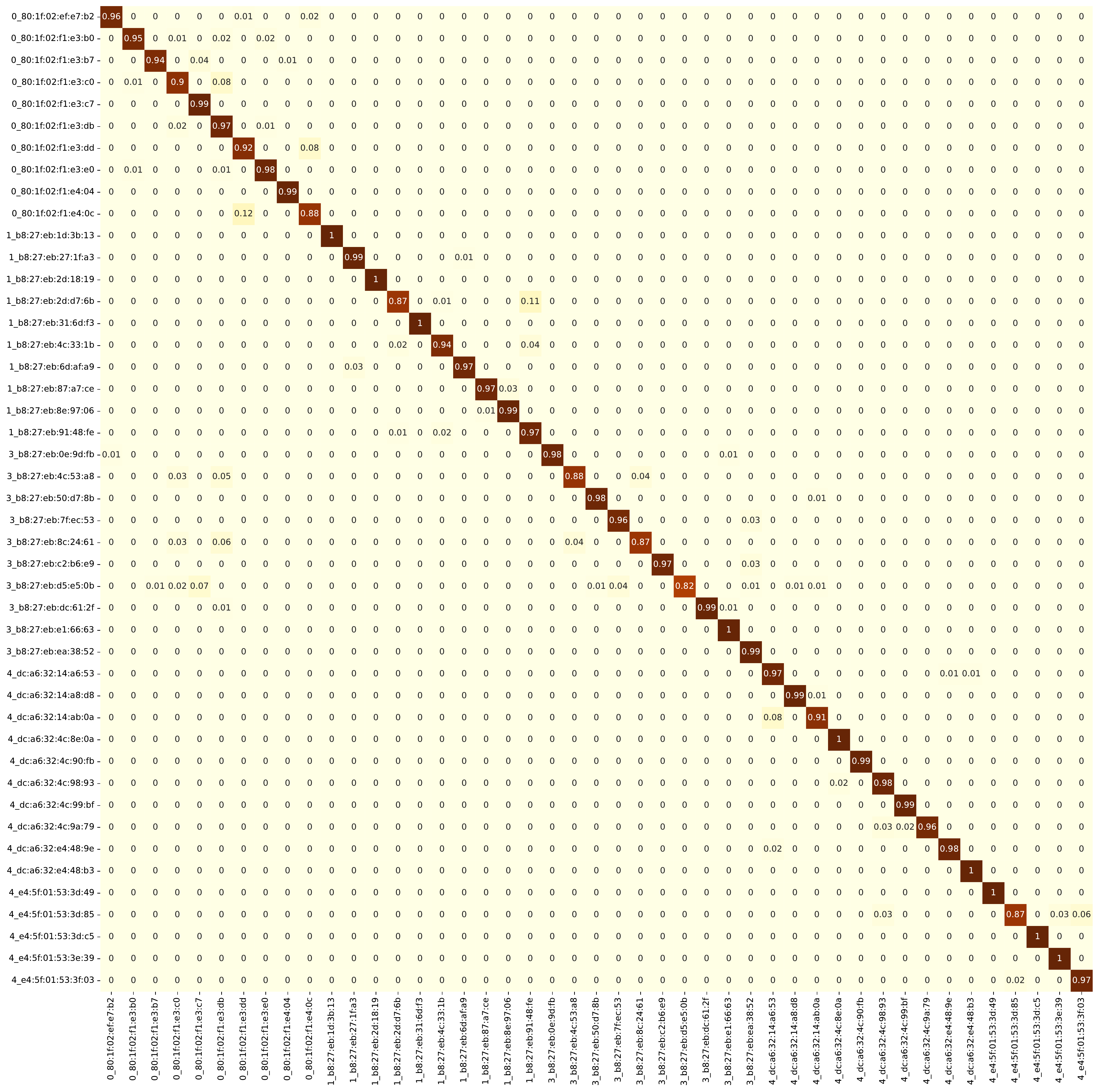}
    \caption{Individual device identification confusion matrix.}
    \label{fig:conf_matrix}
\end{figure*}

It can be seen that the LSTM-1DCNN model is the classification model with the best classification performance, achieving around 0.96 in all the reported metrics in the case of the usage of raw data features. It also shows how the time series approaches using DL-based models are the ones with the best performance, achieving +0.93 in all the reported metrics and improving XGBoost, which was the model with the best performance in similar literature solutions. Besides, \figurename~\ref{fig:conf_matrix} shows the confusion matrix for each device. It can be appreciated that all devices show +0.80 TPR (True Positive Rate), therefore having positive identification of all of them.

From this experiment, it is interesting to observe that the use of raw data features instead of sliding-window ones achieved better results in most of the classification models, something that contrasts with previous works in the literature \cite{Sanchez2021methodology}. This can be a consequence of using a larger dataset than the ones used previously, which also includes information about memory and storage (\cite{Sanchez2021methodology} only included features regarding CPU and GPU).

\section{Adversarial attacks}
\label{sec:attacks}

This section shows the results of the different adversarial attacks tested on the previous ML/DL-based device identification model. The objective is to measure how vulnerable the model is to these attacks if an adversary wants to impersonate a given device or disrupt the identification process.

\subsection{Identification disruption attack}

The objective of an identification disruption attack is to deny the identification of a legitimate device by performing a context-based attack. Here, the adversary seeks to modify the device hardware performance metrics by changing the device environmental and contextual conditions. Then, the device can no longer be identified properly, and the service is affected. This experiment evaluates how resilient the identification solution is to context and environmental changes.

Based on the results in previous research \cite{Sanchez2021methodology,laor2022drawnapart}, the dataset was collected considering context stability conditions as mentioned in Section~\ref{sec:identification}. These conditions ensured that the collected data was not affected by other processes in the device. Therefore context-based attacks leveraging factors such as device rebooting or kernel interruption from other processes are not successful because these were already considered during data collection. Moreover, \cite{Sanchez2021methodology} proved that if the stability and isolation measurements are not included during data collection, the hardware-based identification becomes unstable and does not work properly when the context changes.

\begin{figure}[htpb!]
    \centering \includegraphics[width=0.9\columnwidth]{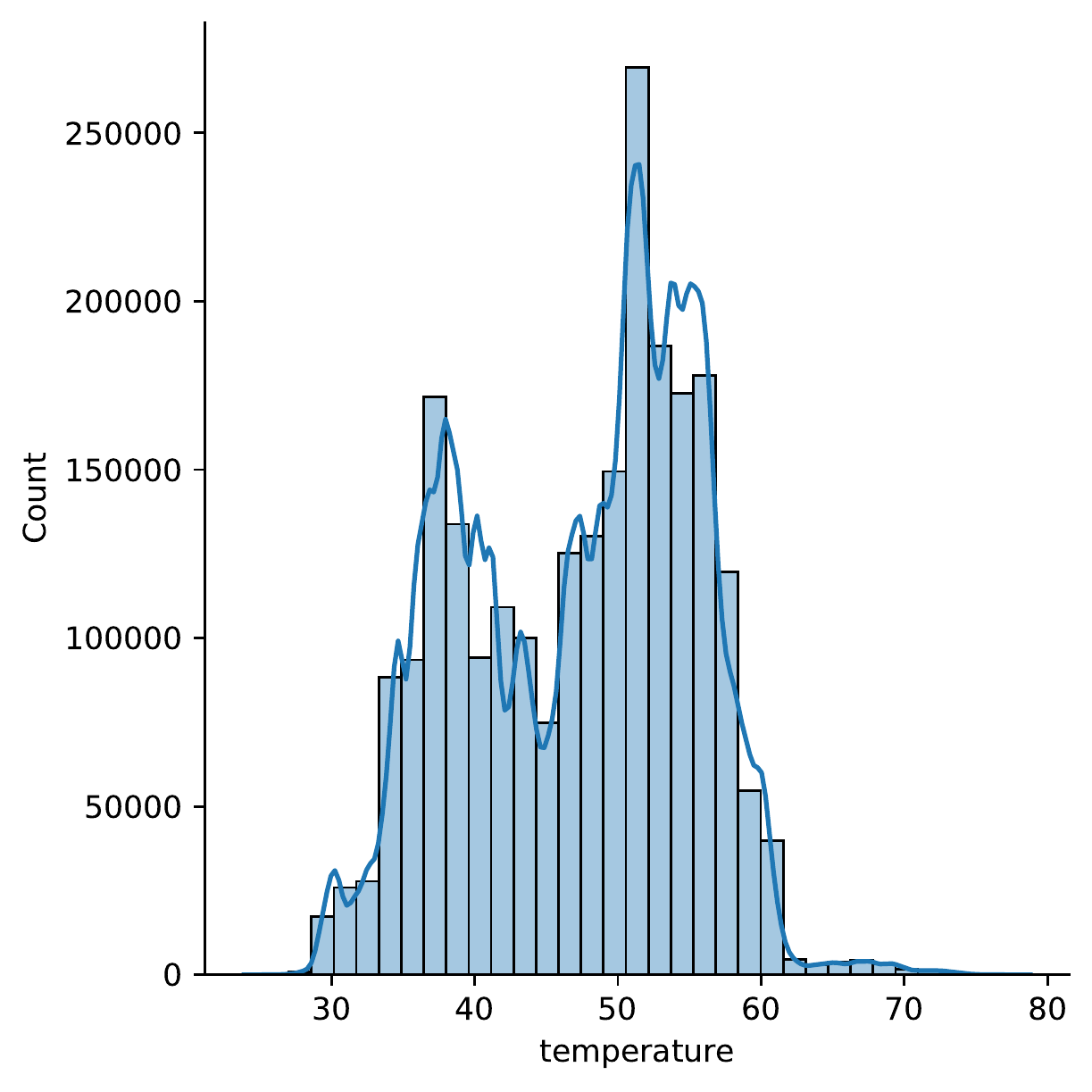}
    \caption{Temperature distribution in the collected dataset.}
    \label{fig:temp}
\end{figure}

Regarding temperature, the environmental conditions were modified by adding heatsinks to the components and turning on/off fans attached to the devices during data collection. In this sense, \figurename~\ref{fig:temp} shows the temperature distribution in the samples contained in the dataset. It can be seen that the temperature during data collection varied from 30ºC to +60ºC. 

The temperature conditions were randomly varied during data collection. Therefore, the train and test datasets employed in Section \ref{sec:identification} do not have a temperature-based bias. However, an attacker might induce new temperature conditions not seen during fingerprint generation to disrupt the identification service. To test this attack, the base dataset of each device was ordered based on the temperature and then divided into train and test samples following an 80/20 ordered split. Then, a new model was generated to compare its performance with the one selected in the previous section. This experiment was repeated both in ascending and descending order. Note that temperature was only used for data ordering and not as a feature.

\tablename~\ref{tab:temp_attack} compares the baseline model with the ones trained by ordering the samples according to the temperature they were collected at. It can be seen how evaluating samples generated at new temperatures does not significantly affect the model performance, with only a 0.03 decrease in the average of the metrics in the case of descending order and even a slight improvement for ascending order. However, the minimum TPR of the evaluated devices (the metric employed for threshold-based identification) is reduced to 0.6682 and 0.6387, respectively. This drop to around 0.65 is observed in two devices in both configurations. Although all the devices can still be identified by setting a threshold in the 0.50 TPR value, these results show that some devices can be more affected by temperature variations.

\begin{table}[htpb!]
	\caption{Temperature-based context attack.}
	\centering
    \scriptsize
    \begin{tabular}{>{\Centering\arraybackslash}p{2cm}>{\Centering\arraybackslash}p{1cm}>{\Centering\arraybackslash}p{1cm}>{\Centering\arraybackslash}p{0.8cm}>{\Centering\arraybackslash}p{0.9cm}>{\Centering\arraybackslash}p{0.7cm}}
        \hline
        \textbf{Model} & \textbf{Accuracy} & \textbf{Avg. Precision} & \textbf{Avg. Recall} & \textbf{Avg. F1-Score} & \textbf{Min. TPR}\\
        \hline
        Baseline order &0.9602 & 0.9626 & 0.9602 & 0.9602 & 0.8045 \\
        \hline
        Temperature ascending order & 0.9621 & 0.9652 & 0.9621 & 0.9619 & 0.6387 \\
        \hline
        Temperature descending order & 0.9394 & 0.9480 & 0.9392 & 0.9402 & 0.6682 \\
        \hline
    \end{tabular}
    \label{tab:temp_attack}
\end{table}{}

This experiment demonstrated that temperature conditions do not excessively impact the device identification performance, as the average performance does not degrade when evaluating data generated under temperatures different from the ones during training. However, for some devices, it can lead to wrong identification if the TPR-based threshold is defined at a high value.

\subsection{Device spoofing attacks}

In the device spoofing attacks, the adversary performs an evasion attack over the already trained ML/DL model, modifying the evaluated data to change the model outputs, so a malicious device is identified as a legitimate one. In this setup, complete knowledge of the model by the attacker was assumed, therefore having a \textit{white-box} evasion attack. 

Besides, as the objective was to fool the device identification model, only targeted attacks make sense to evaluate how easy it is to impersonate other devices. In this sense, one device from each RPi model present in the dataset is selected as the "target class" (constant for attack hyperparameter optimization). Then, the samples from the rest of the devices from each model are used to impersonate that device. Concretely, the selected targets are the RPiZero with MAC \textit{80:1f:02:f1:e3:e0}, the RPi1 with MAC \textit{b8:27:eb:87:a7:ce}, the RPi3 with MAC \textit{b8:27:eb:dc:61:2f}, and the RPi4 with MAC \textit{dc:a6:32:e4:48:9e}.

For the implementation of the attacks, the Adversarial Robustness Toolbox (ART) \cite{art2018} is employed, as it provides straightforward implementations for the attacks detailed in Section \ref{sec:related}. Attack Success Rate (ASR) was employed as the metric for the experiments. In targeted attacks, this metric can be defined as the accuracy of the adversarial samples on the malicious labels. Besides, as the use case was related to device identification using performance-based metrics, the distance between benign and adversarial samples was irrelevant. Note that this metric would be important in other use cases, such as image recognition, where the adversarial and benign samples should not be distinguishable by a human.

The attacks selected to be tested are the ones explained in Section~\ref{sec:related}. However, the DeepFool attack is discarded as it is only untargeted. For this reason, a variant called NewtonFool \cite{jang2017objective} is used in this work. For each attack, an iteration in its main hyperparameters has been performed to find the most successful configuration (the one with a higher ASR). \tablename~\ref{tab:attack_results} shows the ASR results for each attack together with the execution time of the adversarial sample generation.

\begin{table}[htpb!]
	\caption{Adversarial attack results.}
	\centering
    \scriptsize
    \begin{tabular}{>{\Centering\arraybackslash}p{3cm}>{\Centering\arraybackslash}p{3cm}>{\Centering\arraybackslash}p{1.5cm}}
    \hline
    \makecell[c]{\textbf{Attack}} & \makecell[c]{\textbf{Attack Success Rate}} & \makecell[c]{\textbf{Time}} \\
    \hline
    FGSM, $\epsilon=0.05$& 0.3056 & 8.79 s \\
    \hline
    BIM, $\epsilon=0.5$ & 0.8823 &  752.64 s \\
    \hline
    MIM, $\epsilon=0.05$ & 0.8537 & 793.97 s \\
    \hline
    PGD, $\epsilon=0.6$& 0.8823 & 748.06 s \\
    \hline
    NewtonFool, $eta=0.1$ & 0.0994 & 1168.94 s \\
    \hline
    C\&W,$L_2$ & 0.1766 & 63734.18 s \\
    \hline
    C\&W,$L_{inf}$ & 0.0834 & 142087.72 s \\
    \hline
    JSMA, $\theta=0.1$ & Fails & - \\
    \hline
    Boundary Attack & 0.2507 & 379232.28 s \\ 
    \hline
    \end{tabular}
    \label{tab:attack_results}
\end{table}{}

Different target devices were also tested with similar results to the reported in \tablename~\ref{tab:attack_results}. Note that the exact results may vary if other devices were selected as the target, but the objective was to measure the model vulnerability to adversarial attacks. In this sense, FGSM, BIM, and MIM attacks show an ASR over 0.85. All these attacks achieve a +0.50 success in all the devices employed as adversaries. Therefore, these attacks would fully compromise an identification solution setting a threshold in the 0.50 TPR. In contrast, the NewtonFool attack cannot generate adversarial samples complex enough to target the selected class and only generates the misclassification of the data used as a base for crafting adversarial samples. This experiment demonstrated how the model was vulnerable to targeted adversarial evasion attacks, with over 0.85 ASR in some cases. These attacks could perform device spoofing if he/she has enough knowledge about the model or enough trial and error evaluations. 

\section{Defense techniques}
\label{sec:defenses}

This section analyzes how defense techniques can improve the model robustness against ML/DL evasion attacks. ART \cite{art2018} was also used to implement the defense techniques for the model-focused attacks, as it includes several model-focused defense techniques. Note that this defense section focuses on device spoofing attacks because the defenses for context-based attacks were already applied during data collection as explained in the previous sections.

\begin{figure}[htpb!]
    \centering \includegraphics[width=\columnwidth]{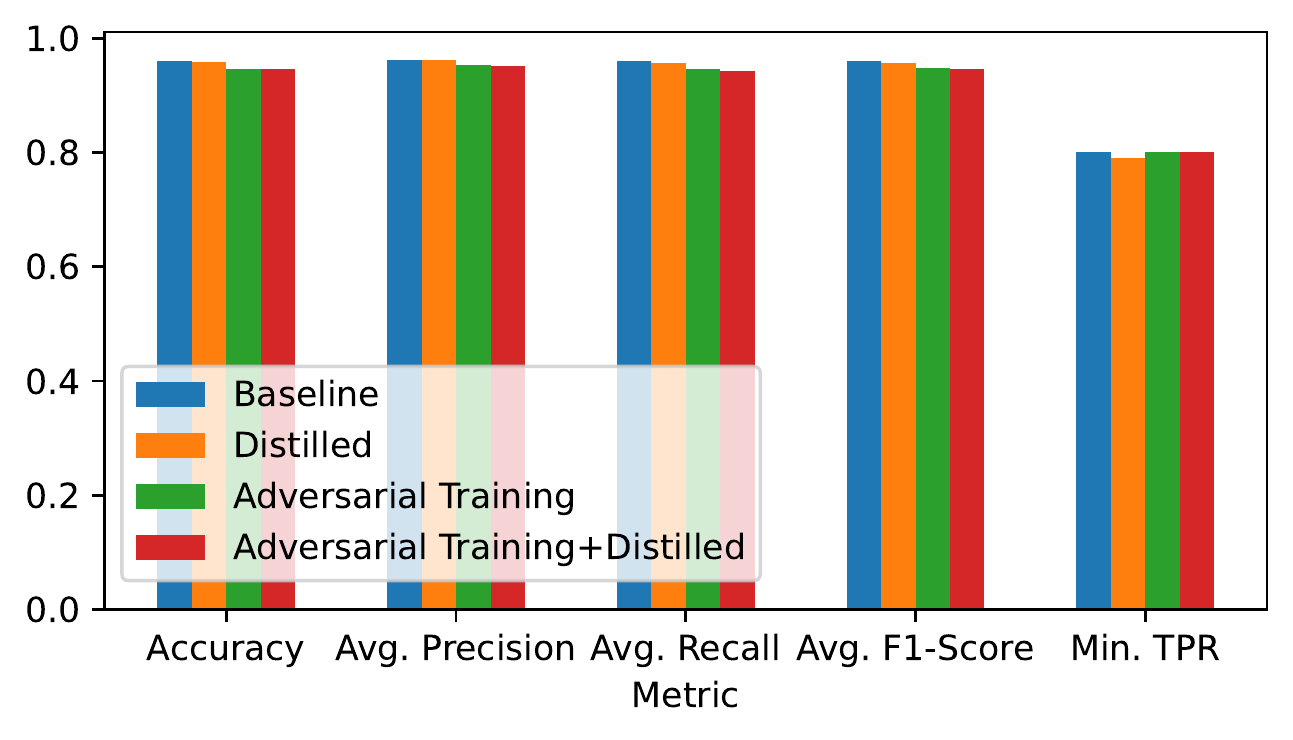}
    \caption{Performance metrics of the robust models.}
    \label{fig:defenses}
\end{figure}

The first defense approach applied was to perform adversarial training, using crafted samples as part of the dataset used for model generation. In this sense, untargeted adversarial samples were generated using FGSM, PGD and BIM attacks and concatenated to the original training dataset. Then, a new model was trained from scratch using the new training dataset. Besides, defensive model distillation \cite{papernot2016distillation} was also applied over the baseline model to compare the robustness of each resulting model. Finally, the model trained using adversarial samples was also distilled. This combination had unstable behavior during model generation, requiring several attempts to avoid gradient explosion issues. \figurename~\ref{fig:defenses} shows the results of each defense model when evaluating the test dataset.

It can be seen how the main performance metrics were not degraded in an impactful manner. Only $\approx$0.02 performance decrease was noticed in accuracy and average precision, recall, and F1-Score. Besides, the minimum TPR was maintained at 0.8 for the adversarial training model and its distilled version. Once it was verified that the robustness techniques did not decrease the identification performance, the next step was to verify if the new models were robust against the evasion attacks. \tablename~\ref{tab:defenses_results} shows the ASR for the different attacks when applied to each model (the baseline one and the ones including robustness techniques).

\begin{table}[htpb!]
	\caption{Attack ASR on the robust models.}
	\centering
    \scriptsize
    \begin{tabular}{>{\Centering\arraybackslash}p{2cm}>{\Centering\arraybackslash}p{1.15cm}>{\Centering\arraybackslash}p{1.15cm}>{\Centering\arraybackslash}p{1.2cm}>{\Centering\arraybackslash}p{1.2cm}}
    \hline
    \multirow{3}{*}{\textbf{Attack}} & \textbf{Baseline Model} & \textbf{Distilled Model} & \textbf{Adversarial Training} & \textbf{Adversarial Training + Distilled} \\
    \hline
    FGSM, $\epsilon=0.05$& 0.3056 & 0.2725 & 0.2704  & 0.1561 \\
    \hline
    BIM, $\epsilon=0.5$ & 0.8823 & 0.3024 & 0.1482 & 0.1631 \\
    \hline
    MIM, $\epsilon=0.05$ & 0.8537 & 0.7950 & 0.1918 & 0.1784 \\
    \hline
    PGD, $\epsilon=0.6$& 0.8823 & 0.2741 & 0.1155 & 0.1235 \\
    \hline
    NewtonFool & 0.0994 & 0.0600 & 0.0846 & 0.0859 \\
    \hline
    C\&W,$L_2$ & 0.1766 & 0.1190 & 0.0953 & 0.0956 \\
    \hline
    C\&W,$L_{inf}$ & 0.0834 & 0.0835 & 0.0848 & 0.0841 \\
    \hline
    JSMA, $\theta=0.1$ & Fails & Fails & Fails & Fails \\
    \hline
    Boundary Attack & 0.2507 & 0.0989 & 0.0886 & 0.0861 \\ 
    \hline,
    \end{tabular}
    \label{tab:defenses_results}
\end{table}{}

In this sense, the model generated by combining adversarial training and distillation was the one offering the best robustness, as the ASR of the different attacks was the lowest in seven of the eight attacks successfully performed. It is worth noting that the ASR of the most impactful attacks in the baseline model, this is BIM, MIM, and PGD, has been decreased from around 0.85/0.88 to 0.12/0.18, which sets the ASR under the 0.5 TPR threshold defined for device identification.

\subsection{Robustness metrics}

There also exist some additional metrics that evaluate how robust a model is by analyzing its parameters and outputs. Therefore, it is relevant to analyze the state-of-the-art metrics in this sense to quantify how the application of robustness techniques improved the model. 

ART \cite{art2018} includes the following metrics regarding model robustness: Cross Lipschitz Extreme Value for nEtwork Robustness (CLEVER) score \cite{weng2018evaluating}, Loss sensitivity \cite{arpit2017closer}, and Empirical robustness \cite{moosavi2016deepfool}. \tablename~\ref{tab:robustness_metrics} shows the values for these metrics using the test dataset as samples for evaluation. Note that CLEVER score is a metric calculated per sample. Therefore, the average and standard deviation are given in the table.

\begin{table}[htpb!]
	\caption{Robustness evaluation results.}
	\centering
    \scriptsize
    \begin{tabular}{>{\Centering\arraybackslash}p{2.15cm}>{\Centering\arraybackslash}p{1.15cm}>{\Centering\arraybackslash}p{1.15cm}>{\Centering\arraybackslash}p{1.15cm}>{\Centering\arraybackslash}p{1.15cm}}
        \hline
        \multirow{3}{*}{\textbf{Metric}} & \textbf{Baseline Model}& \textbf{Distilled Model} & \textbf{Adversarial Training} & \textbf{Adversarial Training + Distilled} \\
        \hline
        CLEVER untargeted, \textit{radius=8, norm=2} & \textit{Avg.}:0.0056 \textit{Dev.}:0.0052 & \textit{Avg.}:0.0058  \textit{Dev.}:0.0059 & \textit{Avg.}:0.0045 \textit{Dev.}:0.0033 & \textit{Avg.}:0.0052  \textit{Dev.}:0.0043 \\
        \hline
        CLEVER targeted, \textit{radius=8, norm=2} & \textit{Avg.}:0.0218 \textit{Dev.}:0.0363 & \textit{Avg.}:0.0239  \textit{Dev.}:0.0305 & \textit{Avg.}:0.0210  \textit{Dev.}:0.0243 & \textit{Avg.}:0.0240  \textit{Dev.}:0.0394 \\
        \hline
        Loss sensitivity & 5.1391 & 4.7920 & 6.8399 & 6.7756 \\
        \hline
        Empirical robustness, \textit{FGSM} $\epsilon=0.05$ & 0.0616 & 0.0613 & 0.0648 & 0.0639\\
        \hline
        \multicolumn{3}{p{4.5cm}}{\textit{Avg.}: Average, \textit{Dev.}: Standard Deviation}\\
    \end{tabular}
    \label{tab:robustness_metrics}
\end{table}{}

Although there is not a very great change on these metrics, and even the results for CLEVER untargeted are worse than the base model, it can be seen how in the case of CLEVER targerted, the score rises 10\% from 0.0218 to $\approx$0.024 in both distilled models. Loss sensitivity score is increased from 5.1391 to 6.8399 and 6.7756 in the adversarial trained and adversarial trained + distilled models, respectively. Finally, Empirical robustness is slightly increased from 0.0616 to 0.0648 and 0.0639, a $\approx$5\%. 

\section{Conclusions and Future Work}
\label{sec:conclusion}

The explosion in IoT device deployment has motivated the development of new device identification solutions based on hardware behavior and ML/DL processing. However, these solutions face adversarial attacks that try to evade their functionality. This work explored the performance of hardware behavior-based device identification. For that, the LwHBench dataset containing samples from 45 Raspberry Pi devices running identical software images was used to train ML/DL classifiers in charge of performing individual identification of each device. A DL model combining LSTM and 1D-CNN layers offered the best performance with an average F1-Score of 0.96, identifying all the devices by setting a threshold in +0.80 TPR. This model improved the performance of previous approaches in the literature. Afterward, a temperature-based attack and nine ML/DL evasion attacks were executed to measure the model performance degradation. In this case, the baseline model was robust against temperature context changes. However, some ML/DL evasion attacks successfully fooled the identification system, reaching up to 0.88 attack success rates and demonstrating its vulnerability to these attacks. Finally, model distillation and adversarial training defense techniques were applied during the model training, improving the model resilience to the ML/DL evasion attacks. These techniques improved the model robustness, being the combination of adversarial training and model distillation the best defense approach. Only a $\approx$0.02 decrease was noticed in accuracy, precision, recall, and F1-Score metrics, without a decrease in the minimum TPR, which is the metric used for setting the threshold for device identification.

In future work, more adversarial attack and defense techniques, such as the ones based on generative models, will be applied to fully improve the solution robustness. Besides, it is planned to add trust metrics in the individual device identification framework, in-depth evaluating the fairness and robustness of the predictions. Another research perspective to be tested is the fully distributed model generation, leveraging federated learning to avoid data sharing and centralization.


\bibliographystyle{IEEEtran}
\bibliography{references}

\begin{IEEEbiography}[{\includegraphics[width=1in,clip]{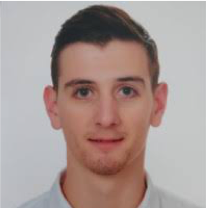}}]{Pedro M. Sánchez Sánchez} is pursuing his PhD in computer science at the University of Murcia. He received the MSc degree in Computer Science from the University of Murcia, Spain. His research interests focus on continuous authentication, networks, 5G, cybersecurity, and machine learning and deep learning.
\end{IEEEbiography}

\begin{IEEEbiography}[{\includegraphics[width=1in,clip]{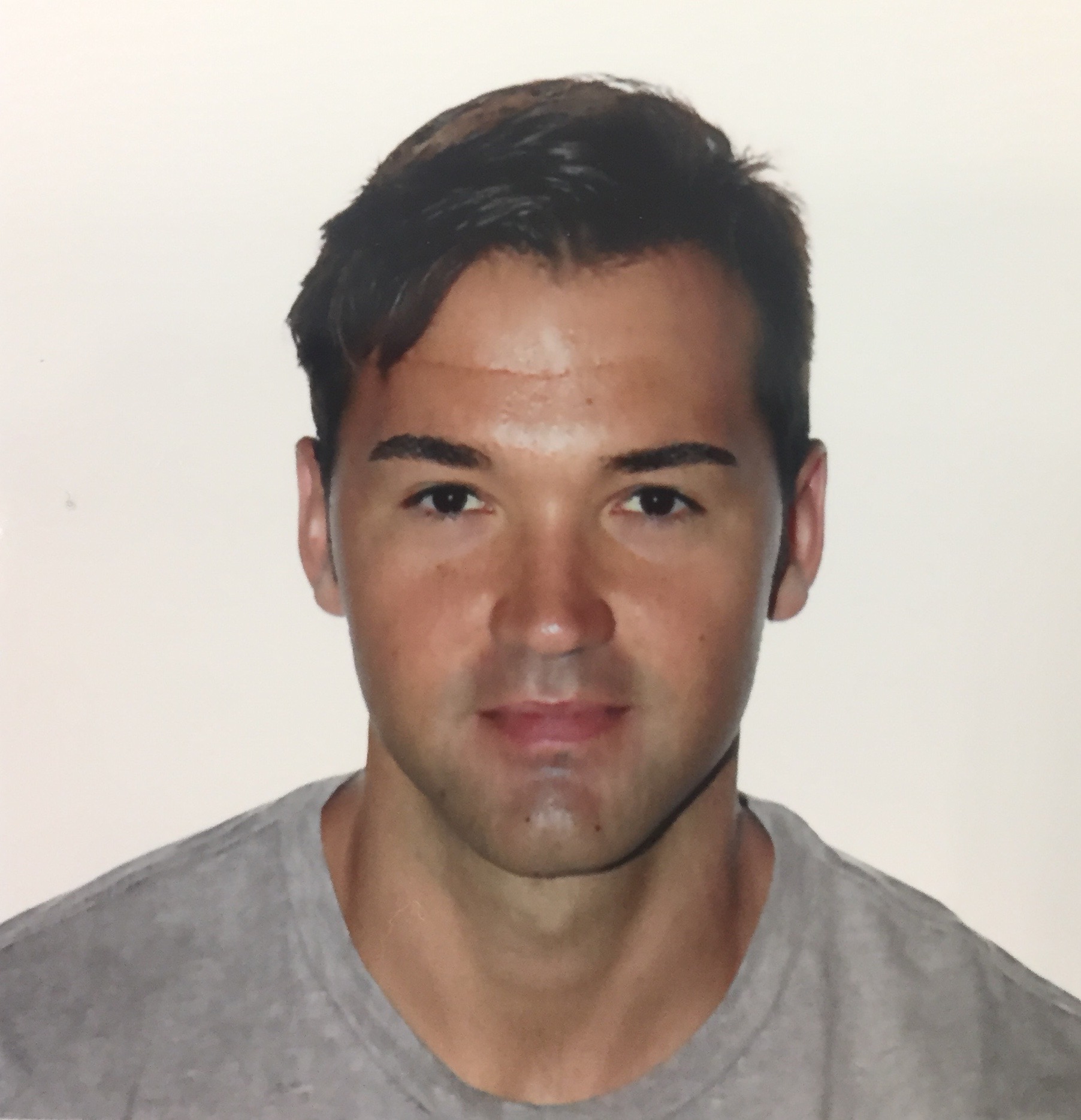}}]{Alberto Huertas Celdrán} is senior researcher at the Communication Systems Group CSG, Department of Informatics IfI, University of Zurich UZH. He received the MSc and PhD degrees in Computer Science from the University of Murcia, Spain. His scientific interests include cybersecurity, machine and deep learning, continuous authentication, and computer networks.
\end{IEEEbiography}

\begin{IEEEbiography}[{\includegraphics[width=1in,clip]{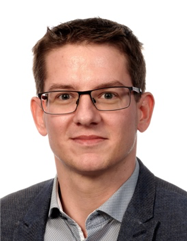}}]{Gérôme Bovet} received his Ph.D. in networks and computer systems from Telecom ParisTech, France, in 2015, and an Executive MBA from the University of Fribourg, Switzerland in 2021. He is the head of data science for the Swiss Department of Defense, where he leads a research team and portfolio of about 30 Cyber-Defence projects. His work focuses on ML and DL approaches, with an emphasis on anomaly detection, adversarial and collaborative learning applied to data gathered by IoT sensors.
\end{IEEEbiography}

\begin{IEEEbiography}[{\includegraphics[width=1in,clip]{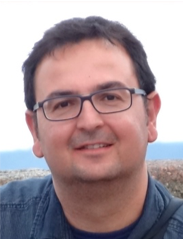}}]{Gregorio Martínez Pérez} is Full Professor in the Department of Information and Communications Engineering of the University of Murcia, Spain. His scientific activity is mainly devoted to cybersecurity and networking. He is working on different national (14 in the last decade) and European IST research projects (11 in the last decade) related to these topics, being Principal Investigator in most of them. He has published 200+ papers in international conference proceedings, magazines and journals.
\end{IEEEbiography}

\end{document}